\documentclass{jpp}
\usepackage{graphicx}

\usepackage[utf8]{inputenc}
\usepackage[T1]{fontenc}
\usepackage{amsmath}
\usepackage{amssymb}
\usepackage{bm}
\usepackage{url}
\usepackage{color}
\newcommand{\beq}{\begin{equation}}
\newcommand{\eeq}{\end{equation}}

\newcommand{\benum}{\begin{enumerate}}
\newcommand{\eenum}{\end{enumerate}}

\newcommand{\abra}[1]{\left\langle{#1}\right\rangle}

\newcommand{\del}{{\bm\nabla}}
\newcommand{\bmA}{{\bm A}}

\newcommand{\bmB}{{\bm B}}
\newcommand{\bmb}{{\bm b}}
\newcommand{\bmE}{{\bm{E}}}

\newcommand{\bmJ}{{\bm J}}

\newcommand{\bmk}{{\bm{k}}}

\def\bmu{\bm{u}}
\newcommand{\bmx}{{\bm{x}}}
\newcommand{\SSSS}{\bm{{\sf S}}}

\newcommand{\xiM}{\xi_\text{M}}
\newcommand{\EEM}{{\cal E}_\text{M}}
\newcommand{\cs}{c_\text{s}}
\newcommand{\kp}{k_\text{peak}}

\newcommand{\Lu}{\text{Lu}}

\newcommand{\Sp}[1]{\text{Sp}\left({#1}\right)}
\newcommand{\intIH}{\mathcal{I}_H(R)}
\newcommand{\cIH}{I_H}
\newcommand{\tcIH}{\tilde I_H}

\usepackage[normalem]{ulem}
\usepackage{xcolor}

\shorttitle{Scaling of the Hosking integral}
\shortauthor{H. Zhou, R. Sharma, and A. Brandenburg}

\title{Scaling of the Hosking integral in decaying magnetically-dominated turbulence}

\author{
Hongzhe Zhou\aff{1,2}\corresp{\email{hongzhe.zhou@su.se}},
Ramkishor Sharma\aff{1,3}
\and
Axel Brandenburg\aff{1,3,4,5}}

\affiliation{
\aff{1}Nordita, KTH Royal Institute of Technology and Stockholm University,
Hannes Alfv\'ens v\"ag 12, SE-10691 Stockholm, Sweden
\aff{2}Tsung-Dao Lee Institute, Shanghai Jiao Tong University,
800 Dongchuan Road, Shanghai 200240,
People's Republic of China
\aff{3}The Oskar Klein Centre, Department of Astronomy,
Stockholm University, AlbaNova, SE-10691 Stockholm, Sweden
\aff{4}McWilliams Center for Cosmology \& Department of Physics,
Carnegie Mellon University, Pittsburgh, PA 15213, USA
\aff{5}School of Natural Sciences and Medicine, Ilia State University,
3-5 Cholokashvili Avenue, 0194 Tbilisi, Georgia
}

\begin{document}

\maketitle

\begin{abstract}
The Saffman helicity invariant of Hosking and Schekochihin (2021, PRX 11,
041005), which we here call the Hosking integral, has emerged
as an important quantity that may govern the decay
properties of magnetically dominated nonhelical turbulence.
Using a range of different computational methods, we confirm that this
quantity is indeed gauge-invariant and nearly perfectly conserved in
the limit of large Lundquist numbers.
For direct numerical simulations with ordinary viscosity and magnetic
diffusivity operators, we find that the solution develops in a nearly
self-similar fashion.
In a diagram quantifying the instantaneous decay coefficients of magnetic
energy and integral scale, we find that the solution evolves along a line
that is indeed suggestive of the governing role of the Hosking integral.
The solution settles near a line in this diagram that is expected for
a self-similar evolution of the magnetic energy spectrum.
The solution will settle in a slightly different position when
the magnetic diffusivity decreases with time, which would be compatible
with the decay being governed by the reconnection time scale rather than
the Alfv\'en time.
\end{abstract}

\section{Introduction}

The subject of decaying turbulence plays important roles in
laboratory and engineering applications \citep{Proudman+Reid54,
Stalp+99}, including superfluid \citep{Nore+97} and supersonic ones
\citep{Kitsionas+09}, as well as those where decaying temperature
fluctuations are of interest \citep{Warhaft+Lumley78}, and in many
areas of astrophysics ranging from star formation \citep{MacLow+98}
to solar physics \citep{Krause+Ruediger75} and especially the early
Universe \citep{Christensson+01, Campanelli+07, Kahniashvili+10, BKT15}.
The application to the early Universe focusses particularly on the decay
of magnetic fields and their associated increase of typical length scales
during the radiation-dominated epoch from microphysical to galactic
scales \citep{Brandenburg+96}; see also \cite{Durrer+Neronov13,subramanian2016} and
\cite{Vachaspati21} for reviews.

The properties of stationary Kolmogorov turbulence are governed by a constant flux of energy to progressively smaller scales.
In decaying turbulence, however, this energy flux is time-dependent.
The rate of energy decay is governed by certain conservation laws,
such as the Loitsiansky integral \citep{Proudman+Reid54} or the Saffman
integral \citep{Saffman67}; see \cite{Davidson00} for a review.
As the energy decreases, the kinetic energy spectrum declines primarily
at large wavenumbers, causing the peak of the spectrum to move toward
smaller wavenumbers or larger length scales.

The presence of a conservation law can also cause an inverse cascade.
A famous example in magnetohydrodynamic (MHD) turbulence is magnetic
helicity conservation, which can lead to a very pronounced increase of
the typical length scale \citep{Frisch+75, PFL76}.
In an alternative approach by \cite{Olesen97}, it has been argued that
the slope of the initial energy spectrum determines the temporal
evolution of the spectrum.
\cite{Olesen97} found the possibility of an inverse cascade for a wide
range of initial spectral slopes.
His argument was based on the observation that the MHD equations are
invariant under rescaling space and time coordinates, $\bmx$ and $t$,
respectively.
Here, the relation between spatial and temporal rescalings depends on
the initial spectrum.
\cite{BK17} found that this relation can, instead, also be governed by
the presence of a conservation law.
In both cases, the formalism of \cite{Olesen97} predicts a self-similar
behavior of the energy spectrum.
This can imply an increase of spectral energy at large length scales,
i.e., an inverse cascade, or at least inverse transfer.\footnote{An
inverse cascade implies a {\em local} transfer in wavenumber space to
smaller $k$.
If the transfer is nonlocal, or if locality in $k$ space
is uncertain, one rather speaks just of inverse {\em transfer}.}

The presence of a conservation law can only affect the behavior of the
system if the conserved quantity is actually finite.
Thus, even though magnetic helicity is always conserved at large magnetic
Reynolds numbers, it may not play a role if the magnetic helicity is zero.
However, MHD turbulence always has nonvanishing fluctuations of magnetic
helicity.
\cite{HoskingSchekochihin2021prx} have shown that the asymptotic
limit, $\cIH$, of the integral of the two-point correlation function of the
local magnetic helicity density $h(\bmx,t)=\bmA\cdot\bmB$,
is invariant in the ideal (non-resistive)
limit and also independent of the gauge $\bmA\to\bmA'=\bmA-\del\Lambda$
for any scalar $\Lambda$.
Here, $\bmB=\del\times\bmA$ is the magnetic field expressed in
terms of the magnetic vector potential $\bmA$.
\cite{HoskingSchekochihin2021prx} argued that the conservation of $\cIH$
determines the decay of non-helical MHD turbulence, and together
with self-similarity, it leads to the inverse cascading in
non-helical magnetically dominated decaying turbulence \citep{BKT15},
which was found independently for relativistic turbulence
\citep{Zrake14}.
\cite{BKT15} argued that the decay was compatible with the conservation
of anastrophy, i.e., the mean squared vector potential, but this explanation remained
problematic owing to the gauge dependence of $\bmA$.
Subsequent work by \cite{Brandenburg+17} with a different initial
condition and lower resolution ($1152^3$ instead of $2304^3$) found a
somewhat faster decay compatible with the conservation of the standard
Saffman integral, which is the two-point correlation function of momentum,
which is proportional to the velocity $\bmu$.
This discrepancy suggests a possible dependence on the magnetic Reynolds
and Lundquist numbers.

\cite{HoskingSchekochihin2021prx} studied the conservation properties of
$\cIH$ using also hyperviscosity and
magnetic hyperdiffusivity to different degrees to study the dependence
on the magnetic Reynolds and Lundquist numbers of the simulation,
as well as justifying the role of the Sweet-Parker regime of
magnetic reconnection.
The latter arises because
topological constraints on the magnetic field prevent ideal relaxation,
as was first suggested by \cite{Zhou+2019,Zhou+2020}, and \cite{Bhat+2021}.
\cite{HoskingSchekochihin2021prx} have shown that the two physical models,
Alfv\'en- vs.\ reconnection-controlled decay, can be unambiguously distinguished by studying
how the energy decay law scales with the hyper-diffusion order, $n$,
and they found the reconnection time scale to be the relevant one.
\cite{Hosking+2022} argue that
it prolongs the effective magnetic decay time and
makes primordial magnetogenesis models consistent with
the bound from GeV observations of blazars \citep{NV10}.

The goal of the present paper is to provide independent evidence for
the conservation properties of $\cIH$
due to magnetic helicity {\em fluctuations}, i.e.,
in the absence of {\em net} magnetic helicity.
We use a range of different methods to assess the reliability of the
results at different stages during the decay.
A particular difficulty is to determine the relevant length scale at
which the Hosking integral is to be evaluated.
We begin by reviewing its definition in section~\ref{SaffmanHelInt},
discuss then our numerical stimulation setup in section~\ref{Setup},
and present our results in section~\ref{Results}.
We conclude in section~\ref{Concl}.

\section{Expressions for the Hosking integral}
\label{SaffmanHelInt}

In this section we introduce different expressions for the Hosking
integral,\footnote{We are grateful to Keith Moffatt for alerting us
to the fact that the formerly used term ``Saffman helicity invariant''
may be misleading, because the term helicity invariant is reserved
for integrals which are chiral in character.
Moreover, Saffman never considered helicity in his papers.
The term ``magnetic helicity density correlation integral'' may be more
appropriate but rather clumsy.
Following \cite{Schekochihin2020}, we now refer to it as the Hosking
integral.
We use the term integral instead of invariant as long as we are not in
the ideal limit.} which we compare later in section~\ref{appx:4methods}.

\subsection{Definition of the Hosking integral}
\label{sec:def_IH}

We recall that the Hosking integral, $\cIH$,
is defined analogously to the Saffman integral in hydrodynamics, and
emerges as the asymptotic limit of the integral of the two-point correlation
function of the local magnetic helicity density
$h(\bmx,t)=\bmA\cdot\bmB$. The latter,
which we call $\intIH$, is given by \citep{HoskingSchekochihin2021prx}
\beq
\intIH=\int_{V_R}\text{d}^3r\ \abra{h(\bmx)h(\bmx+\bm r)}.
\label{eqn:Iv_def1}
\eeq
Here, the angle brackets denote an ensemble average, and $V_R$ is some volume with length scale $R$.
We define the length scale of magnetic fluctuations as
\beq
\xiM=\frac{\int\text{d}k\ k^{-1}E_\text{M}(k)}
{\int\text{d}k\ E_\text{M}(k)},
\label{eqn:xiM}
\eeq
where $E_\text{M}(k)$ is the magnetic energy spectrum.
An alternative length scale can be computed from the spectrum
of the magnetic helicity density, $\xi_h$.
A comparison between $\xiM$ and $\xi_h$ is made in
appendix~\ref{appx:xi}, where we show that, although the ratio
$\xiM/\xi_h$ is not exactly a constant in time, it evolves much more slowly
than the magnetic energy and is of order unity.\footnote{We
thank the anonymous referee for suggesting this.}

When $R$ is much smaller than $\xiM$, we have
\beq
\mathcal{I}_H(R\ll\xiM)\simeq
\int_{V_R}\text{d}^3r\ \abra{h(\bmx)h(\bmx)}\propto R^3,
\eeq
whereas when $R\gg \xiM$, but at the same time much smaller than the system scale $L$,
$\intIH$ reaches a constant asymptotic value, $\cIH$, independent of $R$.
Assuming that the volume average approximates the ensemble average
\citep{HoskingSchekochihin2021prx}, we have
\beq
\cIH=\mathcal{I}_H(\xiM\ll R\ll L)=\frac{1}{V_R}\abra{H_{V_R}^2},\quad
H_{V_R}=\int_{V_R}\text{d}^3r\ h(\bm r).
\label{eqn:Iv_def2}
\eeq
Within this asymptotic range, $\intIH$ is finite and independent
of $R$, because the variance of the magnetic helicity $H_{V_R}$
contained in $V_R$ is expected to scale like
$\abra{H_{V_R}^2}\propto V_R^2\left(V_R/\xiM^3\right)^{-1}\propto V_R$
\citep{HoskingSchekochihin2021prx}.

\subsection{The box-counting method}
To evaluate equation~\eqref{eqn:Iv_def2} in numerical simulations,
we again replace the ensemble average by a moving average, and hence calculate
\beq
\intIH=\frac{1}{VV_R}\int_V\text{d}^3x\ 
\left[\int_{V_R} \text{d}^3r\ h(\bmx+\bm r)\right]^2,
\label{eqn:IH_bc_general}
\eeq
where $V$ is the volume of the simulation box.
In analogy to the algorithm for calculating the Hausdorff dimension,
we call this the box-counting (BC) method.
Upon a Fourier transformation,\footnote{We
define the Fourier transform of $f(x)$ as
$\tilde f(\bmk)=\int\text{d}^3x\,f(\bmx)\,e^{-i\bmk\cdot\bmx}$
and the inverse transformations as
$f(\bmx)=\int(2\pi)^{-3}\text{d}^3k\ 
\tilde f(\bmk)e^{i\bmk\cdot\bmx}$.
}
equation~\eqref{eqn:IH_bc_general} can be recast as a weighted integral
\beq
\intIH=\frac{1}{V}\int\frac{\text{d}^3k}{(2\pi)^3}\ w_R(\bmk)\ h^*(\bmk)h(\bmk).
\label{eqn:Iv_general}
\eeq
The weight function $w_R(\bmk)$ depends on the shape of $V_R$.
For a cubic region with length $2R$, we have
\beq
w_R(\bmk)=w_\text{cube}^\text{BC}(\bmk)\equiv8R^3\prod_{i=1}^3
j_0^2(k_iR),
\eeq
whereas for a spherical region with radius $R$,
\beq
w_R(\bmk)=w_\text{sph}^\text{BC}(k)\equiv\frac{4\pi R^3}{3}
\left[\frac{6j_1(kR)}{kR}\right]^2.
\label{eq:wsph}
\eeq
Here $j_0(x)=\sin x/x$ and $j_1(x)=(\sin x-x\cos x)/x^2$ are the first
and second order spherical Bessel functions, respectively.
Note also that in equation~\eqref{eq:wsph}, $w_\text{sph}^\text{BC}(k)$
depends only on $k=|\bmk|$.
This allows us to rewrite equation~\eqref{eqn:Iv_general} as just a
one-dimensional integral,
\beq
\intIH=\int_0^\infty\text{d}k\ w_\text{sph}^\text{BC}(k)\ \Sp{h},
\label{eq:IHsph}
\eeq
where
\beq
\Sp{h}=\frac{1}{V}\frac{k^2}{(2\pi)^3}\int_{|\bmk|=k}\text{d}\Omega_k \ 
\tilde h^*(\bmk)\tilde h(\bmk)
\label{eqn:Sph}
\eeq
is the Fourier spectrum of the magnetic helicity density and $\Omega_k$
is the solid angle in Fourier space, normalized such that 
$\int\text{d}k\ \Sp{h}=\abra{h^2}$.\footnote{Note
that $\Sp{h}$ is same as $\Theta$ in equation~(32) of
\cite{HoskingSchekochihin2021prx}.}
The magnetic and kinetic energy spectra can then be written analogously
as $E_\text{M}(k)=\Sp{\bmB}/2\mu_0$ and $E_\text{K}(k)=\rho_0\,\Sp{\bmu}/2$,
where $\mu_0$ is the magnetic permeability and $\rho_0$ is the mean density.

\subsection{The correlation integral method}

As an alternative to the BC method, equation~\eqref{eqn:Iv_def1} can be
computed straightforwardly by approximating the ensemble average as a
volume average, i.e.,
\beq
\abra{h(\bmx)h(\bmx+\bm r)}=\frac{1}{V}\int_V\text{d}^3x\ h(\bmx)h(\bmx+\bm r)=
\frac{1}{V}\int\frac{\text{d}^3k}{(2\pi)^3}\ \tilde h^*(\bmk)\tilde h(\bmk)e^{i\bmk\cdot\bm r},
\eeq
which we call the correlation-integral (CI) method.
$\intIH$ can then again be recast in the form of equation~\eqref{eqn:Iv_general},
but the weight functions are now slightly different.
For a cubic region $V_R$ we have
\beq
w_R(\bmk)=w_\text{cube}^\text{CI}(\bmk)\equiv8R^3\prod_{i=1}^3
j_0(k_i R),
\label{eqn:w_spc}
\eeq
and for a spherical region we have
\beq
w_R(\bmk)=w_\text{sph}^\text{CI}(k)\equiv\frac{4\pi R^3}{3}
\frac{6j_1(kR)}{kR}.
\eeq
Note that $w_\text{sph}^\text{CI}(k)$ can also be used in equation~\eqref{eq:IHsph}.

\subsection{The fitting method}
A third way of obtaining $\cIH$ is by extracting the coefficient of the
leading order term of the spectrum of $h$.
At small
$k\ll \xiM^{-1}$ \citep{HoskingSchekochihin2021prx},
\beq
\Sp{h}=\frac{\cIH}{2\pi^2}k^2+\mathcal{O}(k^4),
\label{eq:Sph}
\eeq
given that $h(\bmk)$ is statistically isotropic, and therefore
\beq
\cIH=\lim_{k\to0}\frac{2\pi^2}{k^2}\Sp{h}.
\label{eqn:I_H_fitting}
\eeq

\section{Application to decaying MHD turbulence simulations}
\label{Setup}

\subsection{Basic equations}

We solve the compressible MHD equations in a cubic domain of size $L^3$
using an isothermal equation of state with constant sound speed $\cs$,
so the gas pressure is $p=\rho\cs^2$, where $\rho$ is the density.
We allow for the possibility of hyperviscous and hyperdiffusive
dissipation of kinetic and magnetic energies and solve the equations
for $\bmA$ in the resistive gauge.
The full set of equations is
\beq
\frac{{\rm D}\ln\rho}{{\rm D}t}=-\del\cdot\bmu,
\label{eqn:Dlnrho}
\eeq
\beq
\frac{{\rm D}\bmu}{{\rm D}t}=-\cs^2\del\ln\rho
+\frac{1}{\rho}\left[\bmJ\times\bmB+\del\cdot
(2\rho\nu_n\nabla^{2(n-1)}\SSSS)\right],
\label{eqn:Du}
\eeq
\beq
\frac{\partial\bmA}{\partial t}=\bmu\times\bmB
+\eta_n\nabla^{2n}\bmA,
\label{eqn:dA}
\eeq
where $\bmJ=\del\times\bmB/\mu_0$ is the current density,
$\nu_n$ is the viscosity, $\eta_n$ is the magnetic diffusivity,
${\sf S}_{ij}=(\partial_i u_j+\partial_j u_i)/2-\delta_{ij}\del\cdot\bmu/3$
are the components of the rate-of-strain tensor $\SSSS$,
and $n$ denotes the degree of hyperviscosity or hyperdiffusivity
with $n=1$ corresponding to ordinary viscous diffusive operators
in equations~\eqref{eqn:Du} and \eqref{eqn:dA}, respectively.
Equation~\eqref{eqn:dA} is here formulated in
the resistive gauge, i.e., the scalar potential is
$\varphi=-\eta_n\nabla^{2(n-1)}\del\cdot\bmA$ in the uncurled
induction equation $\partial\bmA/\partial t=-\bmE-\del\varphi$, where
$\bmE=-\bmu\times\bmB+\eta_n\nabla^{2(n-1)}\mu_0\bmJ$ is the electric field.

In some cases, we assume $\nu_n$ and $\eta_n$ to be time-dependent,
which is indicated by writing $\nu_n(t)$ and $\eta_n(t)$, respectively.
In those cases, we assume a power law variation (for $t>\tau$),
\beq
\nu_n(t)=\nu_n^{(0)}[\max(1,t/\tau)]^{r},\quad
\eta_n(t)=\eta_n^{(0)}[\max(1,t/\tau)]^{r},
\eeq
where $r$ is an exponent, $\tau$ is a decay time scale, and $\nu_n^{(0)}$
and $\eta_n^{(0)}$ denote the coefficients at early times.
The motivation for time-dependent $\nu_n(t)$ and $\eta_n(t)$ is two-fold.
On the one hand, perfect self-similarity can only be expected if
the value of $r$ is suitably adjusted \citep{Yousef+04}.
On the other hand, negative values of $r$ are convenient from a
numerical point of view because they allow us to consider Lundquist
numbers that gradually increase as the energy of the turbulence at the
highest wavenumbers decays.

\subsection{Parameters and diagnostic quantities}
\label{Diagnostic}

In the following, we normalize the magnetic energy spectrum
$E_{\rm M}(k,t)\equiv\Sp{\bmB}/2\mu_0$ such that
$\int\text{d}k\ E_{\rm M}(k,t)=\abra{\bmB^2/2\mu_0}\equiv\EEM$,
where $\EEM$ denotes the mean magnetic energy density.
The magnetic integral scale $\xiM$ has been defined in
equation~\eqref{eqn:xiM},
where 
$\xiM^{-1}(t)$ corresponds approximately to the location where
the spectrum peaks, and therefore 
$\xiM^{-1}(0)\approx\kp$.
We define the generalized Lundquist number for hyperdiffusion as
\beq
\Lu_n(t)=\frac{v_\text{A}^\text{rms}\xiM^{2n-1}}{\eta_n},
\eeq
where $v_\text{A}$ is the Alfv\'en velocity.

To characterize the decay of $\EEM$ and $\cIH$, and the increase of $\xiM$,
we define the instantaneous scaling coefficients
\beq
p(t)=-\frac{\text{d}\ln\EEM}{\text{d}\ln t},\quad
p_H(t)=-\frac{\text{d}\ln \cIH}{\text{d}\ln t},\quad
q(t)=\frac{\text{d}\ln\xiM}{\text{d}\ln t}.
\label{eqn:pq_def}
\eeq
Parametric representations of $p(t)$ vs $q(t)$ are useful in
distinguishing different decay behaviors \citep{BK17}.
For purely hydrodynamic turbulence, for example, $p(t)$ and $q(t)$ tend
to evolve along a line $p/q\approx5$ toward a point where $p=10/7$
and $q=2/7$ \citep{Proudman+Reid54}, while for fully helical MHD
turbulence, one sees an evolution along $p/q\approx1$ toward $p=q=2/3$
\citep{Hat84}.
In those cases, the spectrum evolves in an approximately self-similar manner
of the form
\beq
E(k,t)=\xi(t)^{-\beta}\phi\big(k\xi(t)\big),
\eeq
where $\phi(\kappa)$ is a universal function, and $\beta$ is an exponent
that describes the gradual decline of the height of the peak.
Self-similarity is not a stringent requirement, and perfect self-similarity
can also not be expected in a numerical simulation owing to the limited
range of scales that can be resolved.
As shown by \cite{Olesen97}, the ideal MHD equations are invariant
under rescaling; see also appendix~\ref{Rescaling}.
This causes additional constraints that will be discussed below.
To understand the expectations following from self-similarity and
invariance under rescaling, we recall the basic relations involving $p$
and $q$ in appendix~\ref{appx:SelfSim}.

\subsection{Role of Alfv\'en and reconnection times}
\label{Alfven+ReconnectionTimes}

As emphasized above, the ratio $p/q$ is determined by the conserved invariant;
see \cite{BK17} and appendix~\ref{appx:SelfSim}.
If $B^2\xiM^{1+\beta}\sim \EEM\xiM^{1+\beta}$ remains constant during the decay,
then $\EEM\xiM^{1+\beta}\propto t^{-p+q(1+\beta)}\propto t^0$,
which gives
\beq
p=(1+\beta)q.
\label{eqn:pq_beta}
\eeq
In particular, $\beta=0$ when magnetic helicity is conserved, and
$\beta=3/2$ when the Hosking invariant is conserved.\footnote{
\cite{HoskingSchekochihin2021prx} introduced the exponent $\alpha$
and wrote the conserved quantity as $B^\alpha\xiM$.
Then, $\alpha=2/(1+\beta)$, which is 2 and 4/5 for $\beta=0$ and 3/2,
respectively.}

The decay time scale yields a second relation between $p$ and $q$.
When the decay time is the Alfv\'en time, we have
$t_\text{decay}\sim\xiM/v_\text{A}^\text{rms}\sim\xiM/\EEM^{1/2}$,\footnote{This
also implies that $\text{d}\EEM/\text{d}t\sim -\EEM/t_\text{decay}\sim
-\EEM^{3/2}/\xiM$, i.e., the energy dissipation is, as expected,
proportional to $(v_\text{A}^\text{rms})^3/\xiM$.} and therefore
\beq
1=q+p/2.
\label{eqn:pq_Alfven}
\eeq
In that case, together with equation~\eqref{eqn:pq_beta}, we get
$\beta=2/q-3$, which is the relation expected
from the invariance under rescaling the MHD equations; see
appendix~\ref{Rescaling}.
Alternatively, if MHD turbulence decay is controlled by slow
(Sweet-Parker) reconnection, which occurs for $\Lu_n^{1/2n}\lesssim10^4$
\citep{Loureiro+05,Loureiro+07}, we have $t_\text{decay}\sim
\Lu_n^{1/2n}\xiM/\EEM^{1/2}$,
which gives $1=-p/4n+(1-1/2n)q-r/2n+q+p/2$,
and therefore
\beq
(8n-2)q+(2n-1)p=4n+2r,
\label{eqn:pq_Lu}
\eeq
where we have also taken into account the time dependence of the
hyperresistivity, $\eta_n\propto t^r$, so that
$\Lu_n\propto\EEM^{1/2}\xiM^{2n-1}t^{-r}\propto t^{-p/2+(2n-1)q-r}$.\footnote{
Note that for $n\to\infty$, equation~\eqref{eqn:pq_Lu} does not yield
equation~\eqref{eqn:pq_Alfven}.
Mathematically, this is because $n$ also enters in the expression for $\Lu_n$ itself.
To see this, one can write instead $\Lu_n^{1/2m}$ to obtain
$(4m+4n-2)q+(2m-1)p=4m+2r$, which recovers
equation~\eqref{eqn:pq_Lu} 
in the physically meaningful case $m=n$, and equation~\eqref{eqn:pq_Alfven}
when $m\to\infty$ with $n<\infty$ enforced.}
Equation~\eqref{eqn:pq_beta} together with equations~\eqref{eqn:pq_Alfven}
or \eqref{eqn:pq_Lu} uniquely determine the values of $p$ and $q$
in terms of $\beta$ or $\beta$, $r$, and $n$, respectively.
For non-helical decaying MHD turbulence, which is proposed by
\cite{HoskingSchekochihin2021prx} to conserve $\cIH$, and therefore
$\beta=3/2$, we have $p=10/9$ and $q=4/9$ with an Alfv\'en time scale
that is independent of $n$ and $r$.
With the reconnection time scale we have instead
\beq
p=\frac{10\,(2n+r)}{26n-9},\ 
q=\frac{4\,(2n+r)}{26n-9}.
\eeq

\subsection{Initial conditions}

As initial condition we use a magnetic field with a given energy spectrum
proportional to $k^4$ for $k<\kp$ and proportional to $k^{-5/3}$ for $k>\kp$,
where $\kp$ denotes the position of the peak of the initial spectrum.
We assume random phases, which make the field Gaussian distributed.
The smallest wavenumber of the domain is $k_1$, and $\kp$ is taken to
be 60 or 200; see Table~\ref{tab:runs} for a summary of the simulations.
The corresponding data files for these runs can be found in the online
material; see \cite{DATA} for the published data sets used to compute
each of the figures of the present paper.

We use the publicly available {\sc Pencil Code} \citep{Pencilcode2021}.
By default, it uses sixth order accurate finite differences and the
third-order Runge-Kutta timestepping scheme of \cite{Williamson80}.
The different methods for calculating $\intIH$ have been implemented and
are publicly available since the revision of May 20, 2022.
By default, the code computes $\bmA$ in the resistive gauge, but we have also implemented the calculation of 
$\bmA_{\rm C}=\bmA-\del\Lambda$ in the Coulomb gauge by solving
$\nabla^2\Lambda=\del\cdot\bmA$ for the gauge potential $\Lambda$.

\begin{table}
\begin{center}
\begin{tabular}{llllllll}
\hline
Run & $N^3$ & $\EEM(0)$ & $k_\text{peak}$ & $\nu_n=\eta_n$ & $n$ & $r$ & $\Lu_n$\\
\hline
K200D3t & $1024^3$ & $4\times 10^{-3}$ & $200$ & $1\times 10^{-14}$ & $3$ & $-3/7$ & $1.2\times 10\to8.5\times 10^4$\\
K200D3c & $1024^3$ & $4\times 10^{-3}$ & $200$ & $1\times 10^{-14}$ & $3$ & $0$ & $1.2\times 10\to1.4\times 10^3$\\
K60D1c & $1024^3$ & $7\times 10^{-3}$ & $60$ & $5\times 10^{-6}$ & $1$ & $0$ & $5.3\times 10\to2.5\times 10^2$\\
K60D1bt & $2048^3$ & $3\times 10^{-1}$ & $60$ & $2\times 10^{-6}$ & $1$ & $-3/7$ & $2.1\times 10^3\to7.5\times 10^3$\\
K60D1bc & $2048^3$ & $3\times 10^{-1}$ & $60$ & $2\times 10^{-6}$ & $1$ & $0$ & $1.3\times 10^3\to4.0\times 10^3$\\
K60D3t & $1024^3$ & $7\times 10^{-3}$ & $60$ & $1\times 10^{-14}$ & $3$ & $-3/7$ & $1.5\times 10^3\to6.3\times 10^6$\\
K60D3bt & $2048^3$ & $3\times 10^{-1}$ & $60$ & $4\times 10^{-16}$ & $3$ & $-3/7$ & $1.6\times 10^5\to3.2\times 10^7$\\
K60D3bc & $2048^3$ & $3\times 10^{-1}$ & $60$ & $4\times 10^{-16}$ & $3$ & $0$ & $1.7\times 10^5\to2.1\times 10^8$\\
\hline
\end{tabular}
\caption{Summary of runs, where $N^3$ is the resolution, $\EEM(0)$ is the initial
magnetic energy density, and the last column lists
the hyper-Lundquist numbers at the beginning
and the end of the simulations.
Runs ending with `c' use a constant (hyper)diffusivity, whereas those with `t' use a time-dependent value $\propto t^{-3/7}$.
}\label{tab:runs}
\end{center}
\end{table}

\section{Results}
\label{Results}

\subsection{$\intIH$ from different methods}
\label{appx:4methods}

\begin{figure}
\centering
\includegraphics[width=0.9\columnwidth]{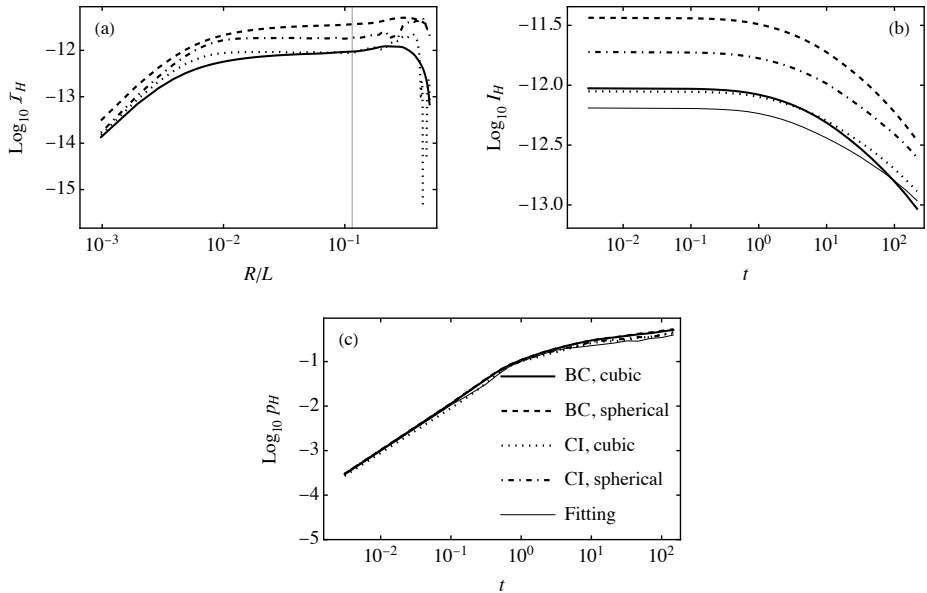}
\caption{Comparing results for run~K60D1c in different methods.
(a) $\intIH$ at $t=0$.
The vertical line indicates $R=0.115L$ with which we compute $\cIH$.
(b) Time evolution of $\cIH$.
(c) Time evolution of the decay exponents $p_H=-\text{d}\ln\cIH/\text{d}\ln t$.
}\label{fig:4methods}
\end{figure}

In Figure~\ref{fig:4methods} we compare for run~K60D1c the temporal
dependence of $\intIH$ and $\cIH$ computed from
the methods introduced in section~\ref{SaffmanHelInt}: BC and CI methods with cubic and spherical
regions $V_R$ for each, and the fitting method.
For the purpose of this comparison, a resolution of $1024^3$ mesh points suffices,
although $\cIH$ is not as well conserved as for higher resolutions.
Our main results are obtained with $2048^3$ mesh points; see Table~\ref{tab:runs}.
The produced $\intIH$ and $\cIH$ curves have different magnitudes, as expected from
the fact that $w_R(\bmk)$ are different; see figures~\ref{fig:4methods}(a) and (b).
However, the scaling properties, i.e., the values of
$p_H=-\text{d}\ln \cIH/\text{d}\ln t$ are unchanged
among all the methods; see figure~\ref{fig:4methods}(c).
In what follows, we use the CI method with cubic regions throughout.

\subsection{Gauge invariance}
\begin{figure}
\centering
\includegraphics[width=0.9\columnwidth]{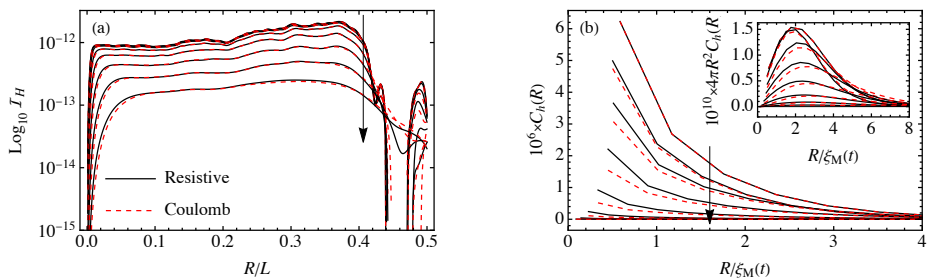}
\caption{Results for run~K60D1c.
(a) Comparing $\intIH$ from different gauges.
(b) The auto-correlation curves $C_h(R)$.
The inset shows $4\pi R^2C_h(R)$.
Note that the abscissa is normalized by $\xiM$, which is
time-dependent.
For both panels, the pairs of curves are taken at $t=0$, $0.2$, $0.5$, $1.5$, $4.6$, $15$, $46$, $147$
in code units from top to bottom, as indicated by the arrows.
}\label{fig:gauge}
\end{figure}

In figure~\ref{fig:gauge}(a) we compare the $R$-dependence of the
correlation integral $\intIH$ at different times under the resistive gauge
(black solid) and the Coulomb gauge (red dashed).
Noticeable differences appear only at later times and at small $R\ll\xiM$,
or when $R$ is close to the system scale $L$.
The differences remain negligible in the asymptotic regime for all $t$.
Thus, even though $h$ itself, and also its spectrum, are gauge-dependent,
$\cIH$ is not \citep{HoskingSchekochihin2021prx}.

The evolution of the auto-correlation $C_h(R)=\abra{h(\bmx)h(\bmx+\bm R)}$
can be computed from $(4\pi R^2)^{-1}\text{d}\intIH/\text{d}R$;
see figure~\ref{fig:gauge}(b).
Using a time-dependent normalization of the abscissa,
we see that $h$ is always correlated at roughly $2\xiM(t)$, as also evident
from the inset.

Let us point out at this point that $\Sp{h}$ is not only gauge
dependent, but it can provide some useful insight into the nature
of gauge dependence.
\cite{Candelaresi+11} used this fact to show that the advective gauge,
where $\varphi=\bmu\cdot\bmA$, can lead to large gradients in the
evolution of $\bmA$, which can cause fatal inaccuracies in the nonlinear
regime.

\subsection{Energy and helicity density spectra}

\begin{figure}
\centering
\includegraphics[width=\columnwidth]{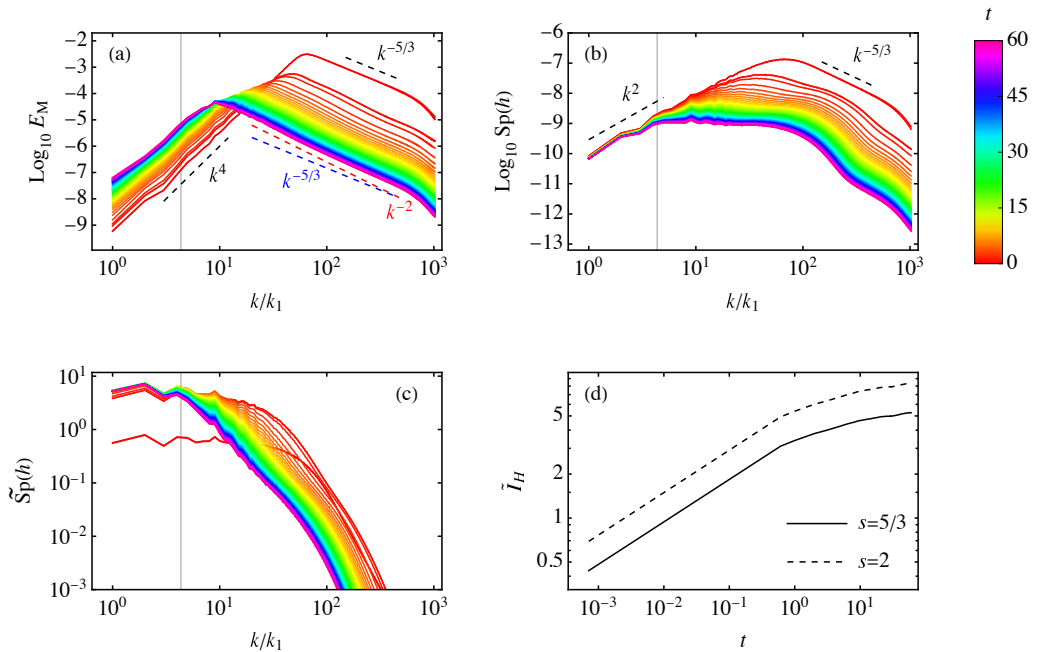}
\caption{Results for run~K60D1bt.
(a) Magnetic energy spectrum.
(b) Spectrum of magnetic helicity density.
(c) The non-dimensionalized and compensated $\Sp{h}$; see equation~\eqref{eqn:Sph_norm}.
(d) Time evolution of the non-dimensionalized $\cIH$, 
equations~\eqref{eqn:IH_norm} and \eqref{eqn:IH_norm_late}.
For the first three panels, the vertical gray lines mark the asymptotic
scale chosen to be $k=2\pi/(2R)=4.35$ with $R=0.115L$, at which value we have computed $\tilde I_H$ in (d).
}\label{fig:spectra}
\end{figure}

In figure~\ref{fig:spectra}, we present magnetic energy spectra
$E_\text{M}(k)\equiv\Sp{\bmB}/2\mu_0$ and magnetic helicity density
spectra $\Sp{h}$ for run~K60D1bt at different times.
From the energy spectra, we see that the inertial range shifts both in
$k$ and in amplitude.
The slope in the inertial range becomes slightly steeper and closer
to $k^{-2}$ at later times.
This is in agreement with previous works \citep{BKT15} and may support the
reconnection-controlled decay picture \citep{Bhat+2021, Zhou+2021}; 
see also \cite{Zhou+2019} for two-dimensional MHD,
and \cite{Zhou+2020} for reduced MHD systems.
The initial $k^4$ subrange, however, does become slightly shallower at
late times when the position of the peak of the spectrum has dropped
below $k/k_1\approx10$.
This is presumably a finite size effect, and so the late time evolution
may not be reliable unless the initial $\kp$ value was large enough and
the total number of mesh points was sufficient to resolve the inertial
and sub-inertial ranges reasonably well.
Since we do not know a priori what are the quantitative requirements
on the initial $\kp$ and on the numerical resolution, we must regard
our results with care and should discuss the possibility of artifacts
as we reach the limits of what can be considered safe.
Thus, we can conclude that, except for very late times,
$\EEM$ has remained nearly self-similar.
The magnetic helicity density spectrum has, just like $E_\text{M}(k)$,
a $k^{-5/3}$ spectrum at the initial time for $k>\kp$.
At small $k$, however, $\Sp{h}$ has a random noise spectrum $\propto k^2$,
which remains unchanged also at late times.
This agrees with our expectations; see equation~\eqref{eq:Sph}, which
allows us to determine $\cIH$ from the spectral value at small $k$.

\subsection{Lundquist-number scaling of decay exponents}

\begin{figure}
\centering
\includegraphics[width=\columnwidth]{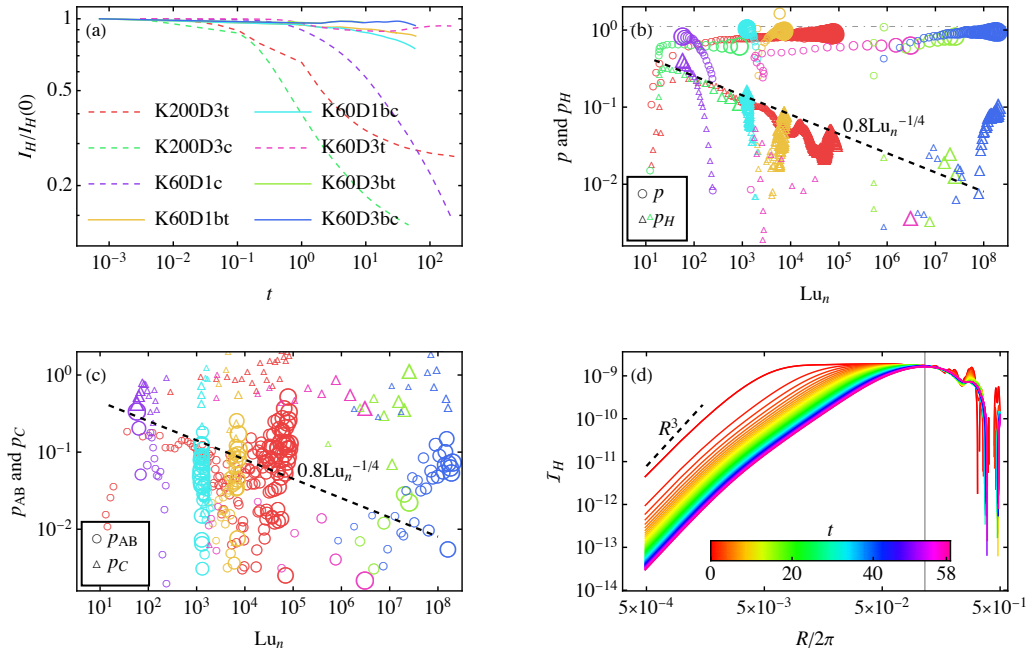}
\caption{(a) Time evolutions of the normalized $\cIH$.
(b) The instantaneous decay exponents of $\cIH$ ($p_H$)
and $\EEM$ ($p$) vs. $\Lu_n$,
and the dash-dotted line indicates $p=10/9$.
The size of the symbols increases with time.
(c) Same as (b), but plotting for the decay exponent of the mean
magnetic helicity ($p_{AB}$), and that of the Hosking cross helicity intetral ($p_C$).
(d) Time evolution of $\intIH$ for run~K60D3bc.
The vertical line indicates the asymptotic scale chosen to be $R/(2\pi)=0.115$.
}\label{fig:Lu}
\end{figure}

In figure~\ref{fig:Lu}(a),
we present the time evolution of the normalized $\cIH$ for all runs with
$R=0.115L$.
With increasing hyper-Lundquist number $\Lu_n$
(cf.\ table~\ref{tab:runs}), $\cIH$ becomes progressively better conserved.
The instantaneous decay exponents $p_H$ vs $\Lu_n$ are plotted in
figure~\ref{fig:Lu}(b) for different runs, along with the energy decay
exponents $p$.
While the latter only has a weak dependence on $\Lu_n$ and approaches
an asymptotic value close to unity
\citep{BKT15,BK17,Reppin+Banerjee17,Bhat+2021,Zhou+2021},
at large $\Lu_n$, $p_H$ decreases and is
found to scale approximately as $\Lu_n^{-1/4}$.

Note that the data points of $p_H$ at the largest $\Lu_n$ values start to level off and even increase with decreasing $\Lu_n$.
We argue that this is an artifact due to the finite size of the computational domain,
and not due to entering the fast-reconnection regime
where the reconnection time scale becomes independent of $\Lu_n$.
In fact, for run~K60D3bc, we have $\Lu_n^{1/n}=20$ at the end of the simulation,
which is still far away from the predicted transition value $\sim 10^4$ \citep{Loureiro+05,Loureiro+07}.

To provide evidence of the limitation from a finite-size box,
in figure~\ref{fig:Lu}(d), we plot the time evolution of the
correlation integral $\intIH$.
At small $R\lesssim\xiM$, we see an $R^3$ scaling.
Thus, $\abra{H_{V_R}^2}/V_R$ is proportional to $V_R\propto R^3$, and
therefore $\abra{H_{V_R}^2}$ is proportional to $V_R^2$, as expected
for a nearly space-filling distribution of magnetic helicity patches of
small scale.\footnote{For this reason,
partially or fully helical magnetic fields would not result in a meaningful
definition of $\cIH$.
In those cases, the conservation of magnetic helicity determines $\beta$.}
At the end of the simulation, $\xiM$ has become comparable to the asymptotic scale
chosen (vertical line), due to which $\cIH$ exhibits an accelerating decay;
see the data points of $p_H$ at the largest $\Lu_n$ values in figure~\ref{fig:Lu}(b).

It is worth noting that the mean magnetic helicity density $|H_V|$ is
never exactly zero, but it is instead several orders of magnitude below
its maximum value.
Interestingly, the decline of the instantaneous decay coefficient of
$|H_V|$, referred to here as $p_{AB}$, follows a similar decline with
$\Lu_n$ as $p_H$; see figure~\ref{fig:Lu}(c).

Following \cite{HoskingSchekochihin2021prx}, we have also performed
a similar analysis for the cross helicity integral, defined as
\beq
\mathcal{I}_c(R)=\int_{V_R}\text{d}^3r\ 
\abra{h_\text{c}(\bmx)h_\text{c}(\bmx+\bm r)},
\eeq
where $h_\text{c}=\bmu\cdot\bmb$ is the cross helicity density.
This is motivated by the fact that the cross helicity is also an ideal
invariant of MHD \citep{Woltjer1958}.
We found that the asymptotic limit, $I_C$, of $\mathcal{I}_C$,
also decays in time, just like $I_H$, but the decay exponent scales
with $\Lu_n$ similarly to that of the magnetic energy density; see figure~\ref{fig:Lu}(c),
where $p_C=-\text{d}\ln I_C/\text{d}\ln t$.
Hence, $I_C$ is not as well conserved as $I_H$.

\subsection{Non-dimensionalizing the Hosking integral}

Since $\cIH$ is conserved, its value can be estimated
from the initial condition.
Through the definition of $h$ we have
\beq
\abra{\tilde h^*(\bmk_1)\tilde h(\bmk_2)}
=\int\frac{\text{d}^3k'}{(2\pi)^3}\frac{\text{d}^3k''}{(2\pi)^3}
\epsilon_{ijk}\epsilon_{abc}k'_k k''_c
\abra{\tilde A^*_i(\bmk_1-\bmk')\tilde A^*_j(\bmk')
\tilde A_a(\bmk_2-\bmk'')\tilde A_b(\bmk'')}.
\label{eqn:hhk}
\eeq
In the Coulomb gauge, we have, for an isotropic field,
\beq
\abra{\tilde A^*_i(\bmk_1)\tilde A_j(\bmk_2)}
=(2\pi)^3\delta^3(\bmk_1-\bmk_2)P_{ij}(\bmk_1)M_A(k_1),
\eeq
where $P_{ij}(\bmk)=\delta_{ij}-k_ik_j/k^2$, and
$M_A$ is related to the magnetic energy spectrum via 
$M_A(k)=2\pi^2 E_\text{M}(k)/k^4$.
Taking advantage of the fact that $\tilde A_i(\bmk)$ is a Gaussian field at $t=0$,
we can decompose the four-point correlations into products of two-point
correlations.
For any wave vectors $\bmk_1$, $\bmk_2$, $\bmk_3$, and $\bmk_4$,
\begin{align}
&\abra{\tilde A^*_i(\bmk_1)\tilde A^*_j(\bmk_2)
\tilde A_a(\bmk_3)\tilde A_b(\bmk_4)}\notag\\
=&\abra{\tilde A^*_i(\bmk_1)\tilde A_j(-\bmk_2)}\abra{
\tilde A^*_a(-\bmk_3)\tilde A_b(\bmk_4)}
+\abra{\tilde A^*_i(\bmk_1)\tilde A_a(\bmk_3)}\abra{
\tilde A^*_j(\bmk_2)\tilde A_b(\bmk_4)}\notag\\
&+\abra{\tilde A^*_i(\bmk_1)\tilde A_b(\bmk_4)}\abra{
\tilde A^*_j(\bmk_2)\tilde A_a(\bmk_3)}\notag\\
=&(2\pi)^6\left[\delta^3(\bmk_1+\bmk_2)\delta^3(\bmk_3+\bmk_4)
P_{ij}(\bmk_1)P_{ab}(\bmk_3)M_A(k_1)M_A(k_3)\right.\notag\\
&+\delta^3(\bmk_1-\bmk_3)\delta^3(\bmk_2-\bmk_4)
P_{ia}(\bmk_1)P_{jb}(\bmk_2)M_A(k_1)M_A(k_2)\notag\\
&\left.+\delta^3(\bmk_1-\bmk_4)\delta^3(\bmk_2-\bmk_3)
P_{ib}(\bmk_1)P_{ja}(\bmk_2)M_A(k_1)M_A(k_2)\right].
\end{align}
Together with equation~\eqref{eqn:Sph}, we find
\beq
\Sp{h}=\frac{k^2}{2}\int_0^\infty\text{d}k'\int_{-1}^1\text{d}\alpha\ 
\frac{k'^2+\mu^2k'^2-2\mu k'|\bmk-\bmk'|}{k'^2|\bmk-\bmk'|^4}
E_\text{M}(|\bmk-\bmk'|)E_\text{M}(k'),
\label{eqn:Sph_gaussian}
\eeq
where $\alpha=\cos(\bmk,\bmk')$, and
\beq
|\bmk-\bmk'|=\sqrt{k^2+k'^2-2\alpha k k'},\ 
\mu=\frac{\alpha k' k-k'^2}
{k'\sqrt{k^2+k'^2-2\alpha k k'}}.
\eeq
Note that equation~\eqref{eqn:Sph_gaussian} is similar to the expression
for $\Sp{\bmB^2}$; see equation~(27) of \cite{Brandenburg+20}.

\begin{figure}
\centering
\includegraphics[width=0.5\columnwidth]{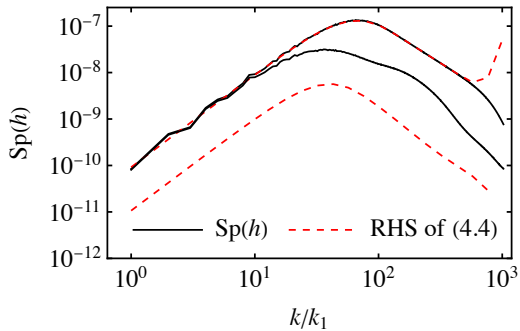}
\caption{For run~K60D1bt, comparing the left-hand (solid black) and right-hand (dashed red) sides of equation~\eqref{eqn:Sph_gaussian},
at two different snapshots, $t=0$ (upper pair) and $t=1$ (bottom pair).}
\label{fig:sph_gaussian}
\end{figure}

Consider a piecewise power-law spectrum,
\beq
E_\text{M}(k)=\left\{\begin{aligned}
& E_\text{peak}\left(k/k_\text{peak}\right)^4, & k\leq k_\text{kpeak} \\
& E_\text{peak}\left(k/k_\text{peak}\right)^{-s}, & k> k_\text{kpeak}
\end{aligned}\right\},
\eeq
where $s=5/3$ and $s=2$ correspond to the
inertial range slopes at early and late times, respectively.
For $s=5/3$ we have the relations
\beq
E_\text{peak}=\frac{20}{17}\EEM\xiM,\quad
k_\text{peak}=\frac{1}{2\xiM}.
\eeq
The leading-order term ($\propto k^2$) of equation~~\eqref{eqn:Sph_gaussian} can be readily obtained by taking $k=0$
in the integrand, which gives
\beq
\Sp{h}^\text{early}=\frac{136 \,E_\text{peak}^2}{95\,k_\text{peak}^3}k^2
+\mathcal{O}(k^3)
=\frac{5120\EEM^2\xiM^5}{323}k^2+\mathcal{O}(k^3).
\label{eqn:Sph_early}
\eeq
For $s=2$ we obtain
\beq
\Sp{h}^\text{late}=\frac{131072\EEM^2\xiM^5}{13125}k^2+\mathcal{O}(k^3).
\label{eqn:Sph_late}
\eeq
In figure~\ref{fig:spectra}(c), we show the compensated spectra normalized using
$s=5/3$,
\beq
\widetilde{\text{Sp}}(h)=\frac{323}{5120\,\EEM^2 \xiM^5k^2}\Sp{h},
\label{eqn:Sph_norm}
\eeq
which is indeed $\sim\mathcal{O}(1)$ initially at small $k$, but then
increases to $\mathcal{O}(10)$ at later times.
Had we used $s=2$, $\widetilde{\text{Sp}}(h)$ only obtains
larger values because the numerical factor in equation~\eqref{eqn:Sph_late}
is smaller than that in equation~\eqref{eqn:Sph_early}.
Indeed, the increase in $\widetilde{\text{Sp}}(h)$ could be caused by
(i) the Lundquist number not being sufficiently high
and/or (ii) the magnetic field becoming non-Gaussian at later times.
To quantify the latter, we find the excess kurtosis of the magnetic field, 
defined as $-3+\sum_{i=1}^3 \abra{B_i^4}/\abra{B_i^2}^2/3$, 
to be $\sim-0.24$ at the end of run~K60D1bt.
This is slightly larger in magnitude compared with earlier work \citep{Brandenburg+20}.
The influence from non-Gaussianity can be explicitly checked by computing the
right-hand side of equation~\eqref{eqn:Sph_gaussian}
and comparing with $\Sp{h}$,\footnote{We thank David Hosking for pointing this out.}
which is shown in figure~\ref{fig:sph_gaussian}.
The Gaussianity is very well satisfied at $t=0$ since we have initialized the field to be so,
but it becomes poor at late times.
In particular, the right-hand side of equation~\eqref{eqn:Sph_gaussian},
which approximates $\Sp{h}$ using $E_\text{M}$, under-estimates the true
value of $\Sp{h}$ by nearly one order of magnitude, although sharing a similar
spectral shape with the latter.
Hence, the non-dimensionalized $\widetilde{\text{Sp}}(h)$ would be larger than unity
also by approximately one order of magnitude, in agreement with figure~\ref{fig:spectra}(c).

Furthermore, we cannot rule out the possibility that the initial non-self-similar
evolution of the magnetic field also contributes to the increase of $\widetilde{\text{Sp}}(h)$.
Since the field is initialized to be Gaussian, local patches of magnetic fields
with the same sign of magnetic helicity are likely not fully helical.
Thus the field configuration close to $t=0$ has an effectively smaller $\xiM$,
and our estimation of $\widetilde{\text{Sp}}(h)$ will be lower at early time.

The magnitude of $\cIH$ can be estimated from equation~\eqref{eqn:I_H_fitting},
and a proper normalization is
\beq
\tcIH=\frac{323}{10240\pi^2\,\EEM^2 \xiM^5}\cIH
\simeq0.0032\,\EEM^{-2} \xiM^{-5}\cIH
\label{eqn:IH_norm}
\eeq
with $s=5/3$ for early times, and 
\beq
\tcIH=\frac{13125}{262144\pi^2\,\EEM^2 \xiM^5}\cIH
\simeq0.0051\,\EEM^{-2} \xiM^{-5}\cIH
\label{eqn:IH_norm_late}
\eeq
with $s=2$ for late times.
Both curves are plotted in figure~\ref{fig:spectra}(d).
The increasing value again highlights the non-Gaussianity of our simulations.

\subsection{Evolution in a $pq$ diagram}

\begin{figure}
\centering
\includegraphics[width=\columnwidth]{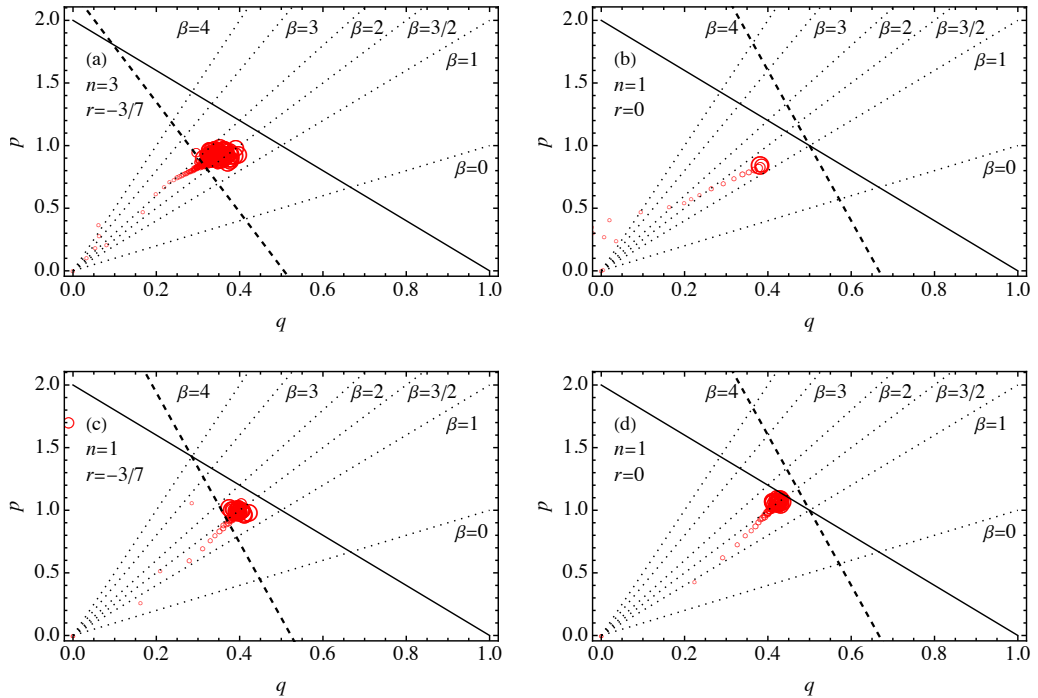}
\caption{Panels (a) to (d) are for 
runs K200D3t, K60D1c, K60D1bt, and K60D1bc, respectively.
The symbol size increases with time.
The dotted, solid, and dashed lines are determined by 
equations~\eqref{eqn:pq_beta}, \eqref{eqn:pq_Alfven},
and \eqref{eqn:pq_Lu}, respectively.}
\label{fig:pq}
\end{figure}

To put our results into perspective, it is instructive to inspect
the evolution in a $pq$ diagram; see section~\ref{Diagnostic}.
In figure \ref{fig:pq}, we plot $pq$ diagrams for four representative runs.
In each panel, the size of the symbols increases with time;
the solid line corresponds to the Alfv\'en relation~\eqref{eqn:pq_Alfven},
and the dashed line gives the reconnection relation~\eqref{eqn:pq_Lu} for each run.

The most reliable runs are those in figures~\ref{fig:pq}(c) and (d),
where $N^3=2048^3$ mesh points have been used.
We clearly see that the solution evolves along the $\beta=3/2$ line,
as expected when the decay is governed by the Hosking integral.
For a constant value of $\eta_1$, figure~\ref{fig:pq}(d), the solution also reaches the
line $p=2(1-q)$, which is referred to as the scale-invariance line.
Note that in some earlier work, it was referred to as the self-similarity
line; see the end of appendix~\ref{RelationToP} for a discussion.
For a time-dependent $\eta_1(t)$, figure~\ref{fig:pq}(c), the solution settles on a point below
the scale-invariance line and is only slightly above the reconnection line.
This might be indicative of reconnection playing indeed a certain role
in the present simulations.

For our hyperviscous run~K200D3t, the solution settles at much smaller
values of $p$ and $q$ and is closer to the reconnection line than to
the scale-invariance line, but the agreement in this case is not very
good either.
How conclusive this is in supporting the idea that reconnection plays
a decisive role must therefore remain open.
Nevertheless, also this solution lies close to the $\beta=3/2$ line,
supporting again the governing role of $\cIH$.

To compare with the case of a fully helical turbulent magnetic
field, we refer to the $pq$ diagram in figure~2(c) of \cite{BK17}.
There, the system evolved along the $\beta=0$ line toward the point
$(p,q)=(2/3,2/3)$.
They also considered the nonhelical case in their figure~2(b), where
the system evolved along the $\beta=1$ line toward the point
$(p,q)=(1,1/2)$.
In a separate study at lower resolution
and with a different initial magnetic field, \cite{Brandenburg+17} 
found an evolution along the $\beta=2$ line toward $(p,q)=(6/5,2/5)$.
Both results were arguably still consistent with an
evolution along $\beta=3/2$ toward $(p,q)=(10/9,4/9)$.

\section{Conclusion}
\label{Concl}

The present results have verified that the Hosking integral is
conserved in the limit of large Lundquist numbers.
This implies that $\cIH$ can indeed control the decay behavior of MHD
turbulence.
On dimensional grounds, one would expect $\beta=3/2$, i.e.,
the solution evolves in a $pq$ diagram along a line where
$p=(1+\beta)q=5q/2$.
Our highest resolution simulations with $2048^3$ mesh points show that
this is compatible with this expectation.

We have also shown that different methods of determining $\cIH$ all lead
to the same result.
The preferred method is based on the magnetic helicity density spectrum,
$\Sp{h}$, which is also the simplest method.
Furthermore, by comparing the resistive and Coulomb gauges, we find that
$\cIH$ is indeed gauge invariant to high precision.

Whether or not the decay time is governed by the reconnection time
rather than the Alfv\'en time remains uncertain, although
our comparison between runs with constant and time-dependent ordinary
magnetic diffusivities, i.e., $\eta_1=\text{const}$ and $\eta_1=\eta_1(t)$,
suggest that the reconnection time scale might indeed be the relevant one.
Clearly, high resolution simulations are required to obtain meaningful
scaling results.
A resolution of $2048^3$ is just beginning to yield conclusive results,
but higher resolution would be desired to address the role of reconnection
more conclusively.

\begin{acknowledgements}
We thank David Hosking, Nuno Loureiro, Kandaswamy Subramanian, and the referees
for helpful comments and discussions.
We also thank Keith Moffatt and Alex Schekochihin for discussions
regarding the problematic naming of the Hosking integral as a
Saffman helicity invariant.
This work was supported by the Swedish Research Council
(Vetenskapsr{\aa}det, 2019-04234).
Nordita is sponsored by Nordforsk.
We acknowledge the allocation of computing resources provided by the
Swedish National Allocations Committee at the Center for Parallel
Computers at the Royal Institute of Technology in Stockholm and
Link\"oping.
\end{acknowledgements}

\appendix
\section{Different definitions of the length scale}
\label{appx:xi}

\begin{figure}
    \centering
    \includegraphics[width=0.5\columnwidth]{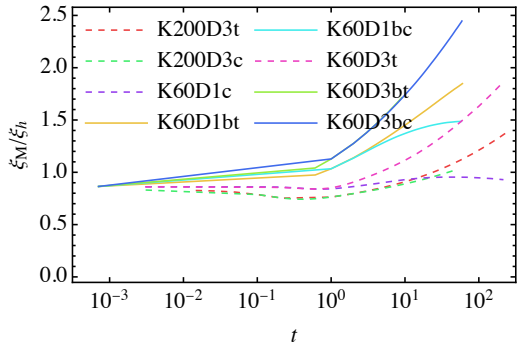}
    \caption{The ratio between the energy integral scale $\xiM$ and the helicity density
    integral scale $\xi_h$.}
    \label{fig:xiM_xih}
\end{figure}

In section~\ref{sec:def_IH}, we used $\xiM$ to characterize 
the length scale of magnetic fluctuations.
An alternative definition can be
\beq
\xi_h\equiv \frac{\int k^{-1}\Sp{h}\ \text{d}k}
{\int \Sp{h}\ \text{d}k}.
\eeq
This is plotted here in figure~\ref{fig:xiM_xih} for all the runs.
Although the ratio is time-dependent, note that this is on a
log-linear scale and thus is a rather weak
dependence in comparison with the
power-law decay of the magnetic energy.

\section{Self-similarity and invariance under rescaling}
\label{appx:SelfSim}

In sections~\ref{Diagnostic} and \ref{Alfven+ReconnectionTimes},
we discussed $pq$ diagrams.
In this appendix, we summarize the essential relations between $p$,
$q$, and the exponent $\beta$, which describes the gradual decline
of the spectral peak as it moves toward smaller $k$.

\subsection{Self-similarity}

A self-similar spectrum implies that its shape is given by a universal
function $\phi=\phi(\kappa)$ such that the spectrum is of the form
\beq
E(k,t)=\xi(t)^{-\beta}\phi\big(k\xi(t)\big),
\label{eq:Ekt}
\eeq
where $\phi(\kappa)$ depends on just one argument
$\kappa\equiv k\xi(t)$, and $\xi=\xi(t)$
is the temporal dependence of the integral scale, which can be used to
describe the gradual shift of the peak of the spectrum towards smaller
$k$.

\subsection{Invariance under rescaling}
\label{Rescaling}

Describing the time dependence of $\xi$ through a power law $\xi\propto
t^q$ means that a rescaling of length, $\bmx\to\bmx'=\bmx\ell$ implies
a rescaling of time, $t\to t'=t\ell^{1/q}$.
Since $E(k,t)$ has dimensions
\beq
[E(k,t)]=[x]^3[t]^{-2},
\eeq
and since $\phi\big(k\xi(t)\big)$ must not change under rescaling, we have
\beq
E(k\ell^{-1},t\ell^{1/q})=\ell^{3-2/q+\beta}\ \left[\xi(t)\ell\right]^{-\beta}\phi(k\xi),
\eeq
and therefore $\beta=2/q-3$ or $q=2/(\beta+3)$; see \cite{Olesen97}.

\subsection{Relation to $p$}
\label{RelationToP}

The exponent $p$ quantifies the temporal scaling of
$\EEM=\int \text{d}k\ E_\text{M}(k,t)\
\propto t^{-p}$.
Inserting \eqref{eq:Ekt}, we have
\beq
t^{-p}\propto\EEM=\xiM(t)^{-(\beta+1)}
\int \text{d}(k\xiM)\ E_\text{M}(k\xiM)
\propto\xiM(t)^{-(\beta+1)}
\propto t^{-(\beta+1)q}
\eeq
and therefore $p=(\beta+1)q$.
Using invariance under rescaling, we have
\beq
p=2(1-q).
\eeq
Table~\ref{tab:coefs} lists various candidates of conserved quantities
and the corresponding values of $q$, $\beta$, and $p$.
In a $pq$ diagram, this represents the line on which the instantaneous
scaling coefficients $p(t)$ and $q(t)$ tend to settle; see \cite{BK17},
where we called this the self-similarity line.
However, if the reconnection time scale really becomes the dominant one,
it may be more meaningful to call $p=2(1-q)$ the scale-invariance line.
It agrees with the assumption of the relevant time scale being the
Alfv\'en time; see equation~\eqref{eqn:pq_Alfven}.
On the other hand, if there really are two distinct scales that evolve
differently, the result cannot be self-similar; see also section 11.2.3 of
\cite{Schekochihin2020} for a discussion.

\begin{table}
\begin{center}
\begin{tabular}{ccccc}
\hline
Conserved quantity & dimension & $q$ & $\beta=2/q-3$ & $p=(\beta+1)q$ \\
\hline
                       &                 &  1  & $-1$ &  0  (=0/4) \\
$\abra{\bmA\cdot\bmB}$ & $[x]^3[t]^{-2}$ & 2/3 &  0   & 2/3 (=4/6) \\
$\abra{\bmA^2}$ (?)    & $[x]^4[t]^{-2}$ & 1/2 &  1   &  1  (=8/8) \\
$\cIH$                  & $[x]^9[t]^{-4}$ & 4/9 & 3/2  & 10/9 \\
$I_u$                  & $[x]^5[t]^{-2}$ & 2/5 &  2   & 6/5 (=12/10) \\
                       &                 & 1/3 &  3   & 4/3 (=16/12) \\
Loitsiansky            & $[x]^7[t]^{-2}$ & 2/7 &  4   & 10/7(=20/14) \\
\hline
\end{tabular}
\caption{
Summary of coefficients.
The question mark on $\abra{\bmA^2}$ indicates that the significance
of this quantity is questionable.
}\label{tab:coefs}
\end{center}
\end{table}

\bibliographystyle{jpp}

\bibliography{refs}

\end{document}